\documentclass[tighten,twocolumn,useAMS,appendixfloats,apj]{likeapj}
\usepackage{etoolbox}
\makeatletter
\patchcmd{\NAT@citex}
  {\@citea\NAT@hyper@{
     \NAT@nmfmt{\NAT@nm}
     \hyper@natlinkbreak{\NAT@aysep\NAT@spacechar}{\@citeb\@extra@b@citeb}
     \NAT@date}}
  {\@citea\NAT@nmfmt{\NAT@nm}
   \NAT@aysep\NAT@spacechar\NAT@hyper@{\NAT@date}}{}{}

\patchcmd{\NAT@citex}
  {\@citea\NAT@hyper@{
     \NAT@nmfmt{\NAT@nm}
     \hyper@natlinkbreak{\NAT@spacechar\NAT@@open\if*#1*\else#1\NAT@spacechar\fi}
       {\@citeb\@extra@b@citeb}
     \NAT@date}}
  {\@citea\NAT@nmfmt{\NAT@nm}
   \NAT@spacechar\NAT@@open\if*#1*\else#1\NAT@spacechar\fi\NAT@hyper@{\NAT@date}}
  {}{}

  \makeatother
\usepackage{mathptmx}

\received{\today}
\revised{}
\accepted{}
\published{}
\submitjournal{\apj}

\shorttitle{}
\shortauthors{R. Serafinelli et al.}

\usepackage{graphicx,natbib,bm,times,amsmath,float}

\hypersetup{linkcolor=blue,citecolor=blue,filecolor=cyan,urlcolor=blue,bookmarksnumbered=true,bookmarksopen=true}
\usepackage[all]{hypcap}

\newcommand{\exi}{\begin{equation}}
\newcommand{\exo}{\end{equation}}

\begin{document}
\title{Unveiling sub-parsec supermassive black hole binary candidates in active galactic nuclei}
\author{Roberto Serafinelli}
\affil{INAF - Osservatorio Astronomico di Brera, Via Brera 28, 20121, Milano, Italy \& Via Bianchi 46, Merate (LC), Italy}
\footnote{\email{roberto.serafinelli@inaf.it}}
\author{Paola Severgnini}
\affil{INAF - Osservatorio Astronomico di Brera, Via Brera 28, 20121, Milano, Italy \& Via Bianchi 46, Merate (LC), Italy}
\author{Valentina Braito}
\affil{INAF - Osservatorio Astronomico di Brera, Via Brera 28, 20121, Milano, Italy \& Via Bianchi 46, Merate (LC), Italy}
\affil{Department of Physics, Institute for Astrophysics and Computational Sciences, The Catholic University of America, Washington, DC, 20064, USA}
\author{Roberto Della Ceca}
\affil{INAF - Osservatorio Astronomico di Brera, Via Brera 28, 20121, Milano, Italy \& Via Bianchi 46, Merate (LC), Italy}
\author{Cristian Vignali}
\affil{Dipartimento di Fisica e Astronomia, Universit\`a degli Studi di Bologna, Via Gobetti 93/2, 40129, Bologna, Italy}
\affil{INAF - Osservatorio di Astrofisica e Scienza dello Spazio di Bologna, Via Gobetti 93/3, 40129, Bologna, Italy}
\author{Filippo Ambrosino}
\affil{INAF - Osservatorio Astronomico di Roma, Via Frascati 33, 00078, Monte Porzio Catone (Roma), Italy}
\affil{INAF - Istituto di Astrofisica e Planetologia Spaziali, Via Fosso del Cavaliere 100, 00133, Roma, Italy}
\affil{Dipartimento di Fisica, ``Sapienza'' Universit\`a di Roma, Piazzale Aldo Moro 5, 00185, Roma, Italy}
\author{Claudia Cicone}
\affil{Institute of Theoretical Astrophysics, University of Oslo, P.O. Box 1029, Blindern, 0315 Oslo, Norway}
\affil{INAF - Osservatorio Astronomico di Brera, Via Brera 28, 20121, Milano, Italy \& Via Bianchi 46, Merate (LC), Italy}
\author{Alessandra Zaino}
\affil{Dipartimento di Matematica e Fisica, Universit\`a degli Studi Roma Tre, Via della Vasca Navale 84, 00146, Roma, Italy}
\author{Massimo Dotti}
\affil{Dipartimento di Fisica ``G. Occhialini'', Universit\`a di Milano-Bicocca, Piazza della Scienza 3, 20126, Milano, Italy}
\affil{INFN - Sezione di Milano-Bicocca, Piazza della Scienza 3, 20126, Milano, Italy}
\author{Alberto Sesana}
\affil{Dipartimento di Fisica ``G. Occhialini'', Universit\`a di Milano-Bicocca, Piazza della Scienza 3, 20126, Milano, Italy}
\author{Vittoria E. Gianolli}
\affil{INAF - Osservatorio Astronomico di Brera, Via Brera 28, 20121, Milano, Italy \& Via Bianchi 46, Merate (LC), Italy}
\affil{Dipartimento di Fisica ``G. Occhialini'', Universit\`a di Milano-Bicocca, Piazza della Scienza 3, 20126, Milano, Italy}
\author{Lucia Ballo}
\affil{XMM-Newton Science Operations Centre, ESAC/ESA, PO Box 78, 28691, Villanueva de la Ca\~nada, Madrid, Spain}
\author{Valentina La Parola}
\affil{INAF - Istituto di Astrofisica Spaziale e Fisica Cosmica di Palermo, Via Ugo La Malfa, 90146, Palermo, Italy}
\author{Gabriele A. Matzeu}
\affil{INAF - Osservatorio Astronomico di Brera, Via Brera 28, 20121, Milano, Italy \& Via Bianchi 46, Merate (LC), Italy}
\affil{European Space Agency (ESA), European Space Astronomy Centre (ESAC), 28691, Villanueva de la Ca\~nada, Madrid, Spain}

\begin{abstract}
Elusive supermassive black hole binaries (SMBHBs) are thought to be the penultimate stage of galaxy mergers, preceding a final coalescence phase. SMBHBs are sources of continuous gravitational waves, possibly detectable by pulsar timing arrays; the identification of candidates could help in performing targeted gravitational wave searches. Due to their origin in the innermost parts of active galactic nuclei (AGN), X-rays are a promising tool to unveil the presence of SMBHBs, by means of either double Fe K$\alpha$ emission lines or periodicity in their light curve. Here we report on a new method to select SMBHBs by means of the presence of a periodic signal in their {\it Swift}-BAT 105-months light curves. Our technique is based on the Fisher's exact g-test and takes into account the possible presence of colored noise. Among the 553 AGN selected for our investigation, only the Seyfert 1.5 Mrk 915 emerged as possible candidate for a SMBHB; from the subsequent analysis of its light curve we find a period $P_0=35\pm2$ months, and the null hypothesis is rejected at the $3.7\sigma$ confidence level. We also present a detailed analysis of the BAT light curve of the only previously X-ray-selected binary candidate source in the literature, the Seyfert 2 galaxy MCG+11-11-032. We find $P_0=26.3\pm0.6$ months, consistent with the one inferred from previously reported double Fe K$\alpha$ emission lines.
\end{abstract}

\keywords{ galaxies: active -- X-rays: galaxies -- galaxies: Seyfert}

\section{Introduction}
\label{sec:intro}

It is now well established that massive galaxies host supermassive black holes (SMBH, $M_\text{BH}\gtrsim10^{6}M_\odot$) in their central regions \citep[e.g.,][]{kormendy13}. Such extreme astrophysical objects are likely linked to the evolution of their host galaxies. In fact, stellar bulge properties such as velocity dispersion \citep[e.g.,][]{ferrarese00} or stellar mass \citep[e.g.,][]{haring04}, are found to correlate with the mass of the central SMBH.\\
\indent In $\Lambda$ cold dark matter cosmological models ($\Lambda$CDM) galaxies grow hierarchically through frequent mergers \citep[e.g.,][]{cole00,volonteri03}. Galaxy mergers have been proposed as one of the main mechanisms by which nuclear activity, star formation or outflows are triggered \citep[e.g.,][]{dimatteo05,hopkins06,dimatteo08}. One of the most interesting possible outcomes of this process is the formation of a supermassive black hole binary (SMBHB) system. During a galaxy merger, if both SMBHs are accreting at kpc-scale distance from each other, the source will appear as a dual active galactic nucleus (AGN) \citep[e.g.,][and references therein]{koss12,mcgurk15,koss18,derosa18,derosa19}. Subsequently, the two SMBHs drift towards the center of mass because of dynamical friction \citep{dosopoulou17} and form an inspiralling binary system at sub-pc-scale separations \citep[e.g.,][]{begelman80}, which will eventually merge \citep[e.g.,][]{dotti12,mayer13,colpi14}. Theoretical models predict that, after the two black holes bind in a binary, dynamical friction becomes inefficient, possibly preventing the binary to merge in a time shorter than the Hubble time \citep[e.g.,][]{milosavljevic03}. However, in presence of nuclear gas inflows \citep[e.g., ][]{dotti12}, or deviations from spherical symmetry in the potential of the nucleus \citep{sesana15}, the binary can still transfer enough energy to its surroundings to merge within $t_\text{coal}\sim1$ Gyr. Despite the expected abundance of such objects, the detection of SMBHBs remains elusive and only limited indirect observational evidence has been found so far \citep[see][for a recent review]{derosa19}. This is mainly due to the angular resolution of current telescopes, which do not have a resolving power below a few parsecs at most redshifts and wavelengths \citep[e.g.,][]{rodriguez06}, insufficient to distinguish emitting binary systems.\\
\indent Hydrodynamic numerical simulations usually assume that a SMBHB excavates a cavity in the surrounding gas, which forms a circumbinary accretion disk \citep[e.g.,][]{noble12,farris15}. At smaller scales, accretion possibly occurs through minidisks around each black hole inside the cavity \citep[e.g.,][]{hayasaki08,dascoli18}. While the circumbinary disk is thought to be responsible for the optical and IR emission \citep{dorazio15a}, the UV and X-rays should be mainly emitted by the inner minidisks \citep{sesana12,dascoli18}. In particular, the UV and X-ray fluxes are expected to be periodic, since the accretion rate of the minidisks is periodically fed by streams of gas flowing from the outer circumbinary disk \citep[e.g.,][]{hayasaki08,haiman09,farris15}, with periods comparable to the binary period. Assuming a Keplerian regime, for SMBHBs with a separation of the order of $10^{-3}$ pc, we expect a periodicity with period longer than about a year.  As these emissions originate very close to each black hole, UV and X-ray periodic modulation \citep[e.g.,][]{sesana12,roedig14,farris15,haiman17} can be used to identify possible SMBHB. Additionally, the two minidisks may produce double-peaked Fe K$\alpha$ emission lines in the X-ray band \citep[e.g.,][]{sesana12,popovic12,mckernan15}, with energies Doppler-shifted by the minidisk relative to orbital motion. So far, the only X-ray-selected SMBHB candidate is the Seyfert 2 galaxy MCG+11-11-032 \citep{severgnini18}. The detection of two peaks (at $4\sigma$ and $2\sigma$, respectively) in the Fe K$\alpha$ energy range, along with the visually identified harmonic behaviour of the 123-month {\it Swift}-Burst Alert Telescope (BAT) X-ray light curve with a period of $P\sim25$ months, strongly suggested the presence of a SMBHB.\\
\indent A periodical variability from the inflow streams is also expected in the optical band, although it may be overwhelmed by the non-periodic emission from the circumbinary disk. Despite the fact that the  UV and X-ray data are likely the best electromagnetic (EM) bands for periodicity searches in AGN, regular monitoring observations in these bands, spanning several years, are rare. For these reasons, the search for EM periodicity has been mainly undertaken in the optical band. \cite{graham15b} found 111 optically-selected SMBHB candidates in the Catalina Real-Time Survey \citep[CRTS,][]{drake09}, and \cite{charisi16} identified 33 further objects in the Palomar Transient Factory \citep[PTF,][]{rau09}. In addition, the presence of a periodic signal has also been reported in single-source studies, such as OJ 287 \citep[e.g.,][]{sillanpaa88}, PG 1302-102 \citep{graham15a}, NGC 5548 \citep{bon16}, and more recently Mrk 231 \citep{kovacevic20}. In particular, OJ 287 shows recurring flares with a precision so high that it is possible to predict the following flare with an accuracy of $\sim4$ hours on a $\sim12$-year period \citep{laine20}. However, a significant fraction of those candidates are likely false positives. In fact, such a large number of SMBHBs should have produced a GW cosmic background above the current pulsar timing array limit (PTA, see \citealp{arzoumanian18} for the latest NANOGrav Collaboration limit, \citealp{lentati15} for the latest European PTA limit, and \citealp{shannon15} for the Parkes PTA limit), while no detection has been reported so far \citep{sesana18}.\\
\indent By  exploiting  the 105-Months {\it Swift}-BAT Survey \citep{oh18}, here we propose a novel method to search for periodicities in colored noise X-ray light curves based on the Fisher's exact g-test \citep{fisher29} and its probability of false alarm due to colored noise fluctuations. A recent analysis of the X-ray light curves of a subsample of $\sim220$ AGN, selected by their large excess variance,\footnote{The excess variance \citep[e.g.,][]{nandra97} is a method to quantify variability, defined as $\sigma_{\rm xs}^2=S^2-\overline{\sigma(x)^2},$ where $S^2$ is the variance of the light curve and $\sigma(x)$ is the photometric error.} of the BAT survey \citep{liu20} did not find any AGN with evidence for periodic behaviour in their X-ray light curves. The search was based on the power spectrum fitting \citep{vaughan05} of colored noise light curves, i.e. those curves whose power spectral densities are not flat, or white (see Sect.~\ref{sec:method} for details).\\ 
\indent The paper is structured as follows: in Section~\ref{sec:toolbox} we review the power spectrum and the X-ray variability tools usually adopted, while in Section~\ref{sec:method} we describe the proposed search method on simulated light curves. In Section~\ref{sec:data} we apply the method to real X-ray light curves in the 105-Months {\it Swift}-BAT Survey and test the selected candidates by using sinusoidal fitting and epoch folding techniques. MCG+11-11-032 is also discussed in Section~\ref{sec:mcg}. Finally, we summarize and discuss our results in Section~\ref{sec:summary}.\\
\indent Throughout the paper, we adopt a flat $\Lambda$CDM cosmology $H_0=70$ km s$^{-1}$ Mpc$^{-1}$, $\Omega_\Lambda=0.7$ and $\Omega_M=0.3$.

\section{Identifying a periodic trend in a stochastic light curve}
\label{sec:toolbox}

\begin{figure*}
\centering
\includegraphics[scale=0.48]{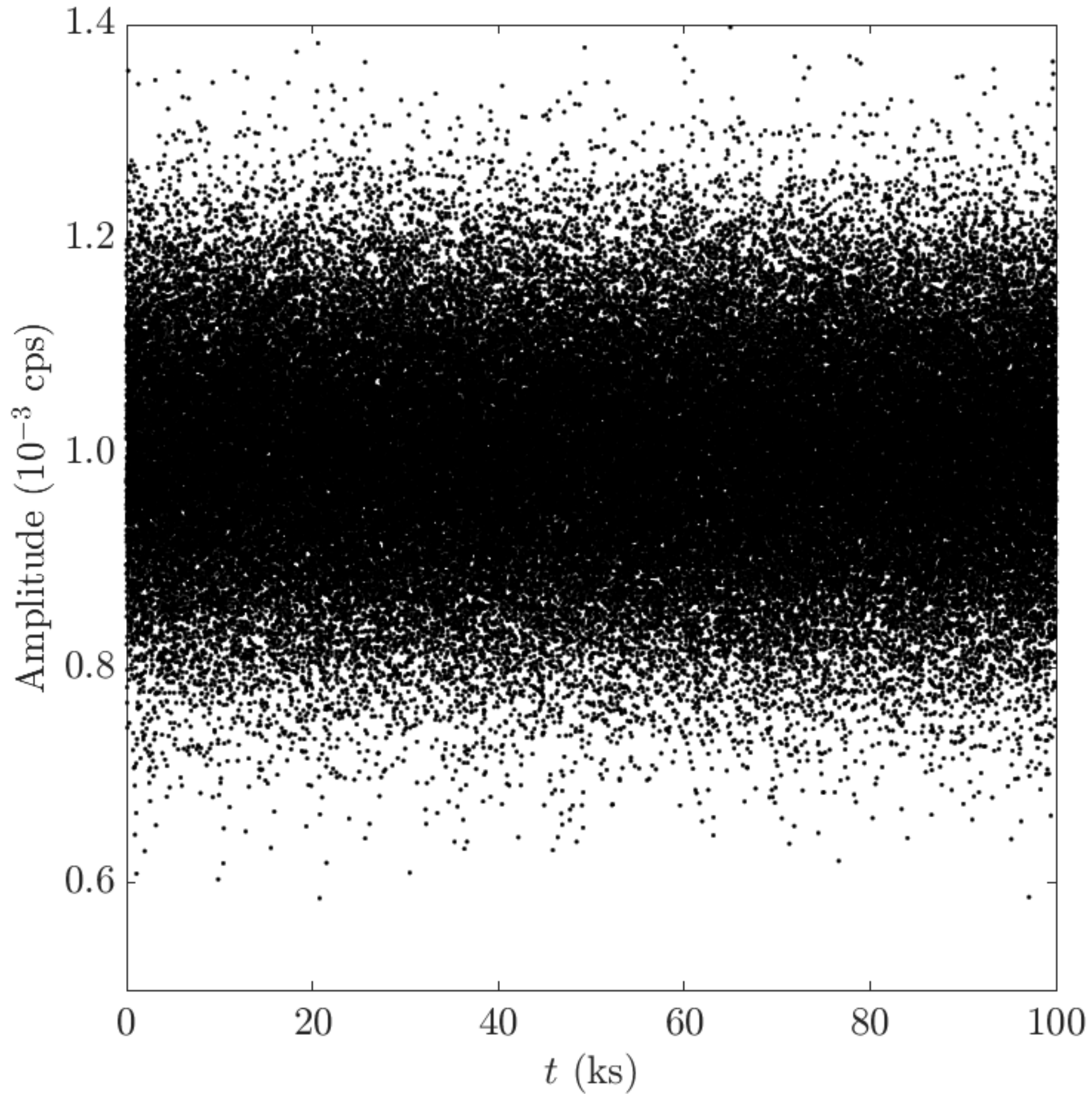}\includegraphics[scale=0.48]{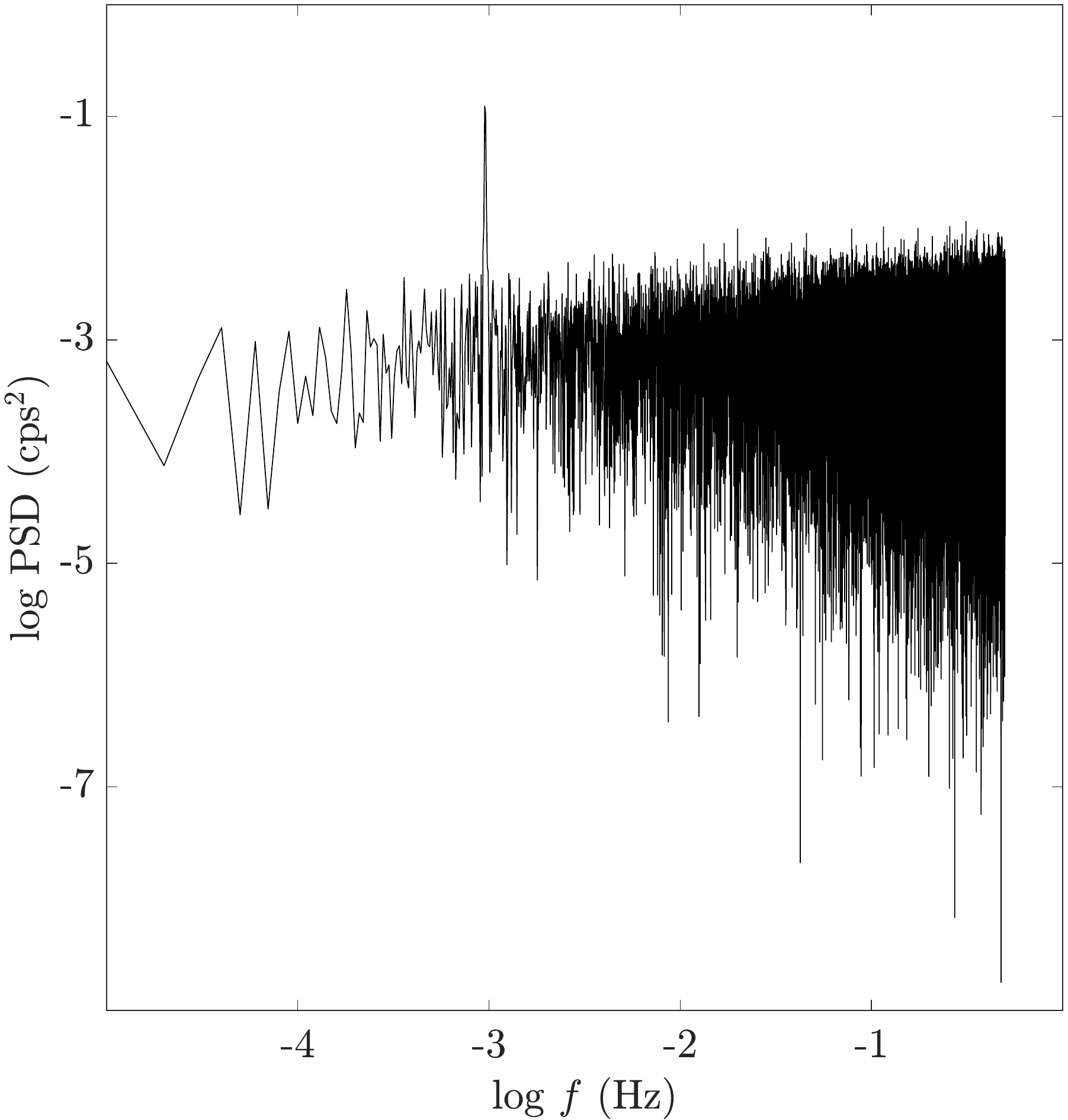}
\caption{{\it Left}. Light curve of $100$ ks simulated data, sampled at $1$ s, produced by the superposition of stochastic white noise and a sinusoid whose amplitude is set to $0.1$ times the standard deviation of the noise and period set to $P_0=1$ ks. No evident periodicity can be inferred. {\it Right}. The periodogram of the same light curve shown in the left panel, showing a significant $\delta$-like peak at $f=10^{-3}$ Hz.}
\label{fig:lcps}
\end{figure*}

The X-ray emission from active galactic nuclei is well known to be very variable \citep[e.g.,][]{mchardy04,markowitz04,vagnetti16,falocco17}, even at hard ($E>10$ keV) energies \citep[e.g.,][]{soldi14}. The X-ray variability of AGN is typically a superposition of stochastic processes and therefore is not easily modeled as a function of time and is usually referred to as noise.\\
\indent Identifying periodic trends in light curves may result in a challenging task, because a sinusoidal or quasi-sinusoidal curve may be hidden in noise and be hardly recognizable in the time domain. A very powerful tool to identify periodicities is the power spectral density (PSD), or power spectrum, which represents the mean value of the squared amplitude as a function of the temporal frequency $f$. The most frequently used tool to represent the power spectrum is the periodogram $P_\text{S}(f)$, which is defined as the squared Fourier transform of the light curve, i.e. $P_\text{S}(f)\propto\left|\hat{X}(f)\right|^{2}=\hat{X}(f)\times\hat{X}^*(f)$, where $\hat{X}$ is the Fourier transform of the light curve $x(t)$ and $\hat{X}^*$ its complex conjugate\footnote{It should be reminded that all real and simulated data are discrete, althought to simplify the notations we will denote them as continuous throughout the whole article}. We subtract the mean value of the light curve $x_0\equiv\left<x(t)\right>$ before computing the Fourier transform, to remove the constant component of the light curve. We also neglect all frequency values larger than or equal to the Nyquist frequency $f_\text{Nyq}=1/2\delta t$, where $\delta t$ is the sampling time, and their corresponding periodogram values, in order to exclude points due to aliasing. Several normalizations may be defined for the periodogram \citep[e.g.,][]{papadakis93,vaughan03,emmanoulopoulos16}, but since we intend to study individual sources one by one (see Sect.~\ref{sec:data}), we choose a unitary normalization, and therefore $P_\text{S}(f)=\left|\hat{X}(f)\right|^2$. This quantity is ideal to find periodicities, given that the response to a pure sinusoid is a Dirac $\delta$-function. PSD is regularly used to identify periodicities in pulsar astronomy at many electromagnetic (EM) bands \citep[e.g.,][]{israel16,ambrosino17,mickaliger18}, and in searches for their continuous gravitational wave (GW) counterparts \citep[e.g.,][]{aasi15}, detection of quasi-periodic oscillations (QPO) in X-ray binaries \citep[e.g.,][]{zhang96}, but also to search for both short- \citep[e.g.,][]{gierlinski08,miniutti19} and long-term periodicities in AGN light curves \citep[e.g.,][]{bon16,liu20}. Moreover, the Fourier transform operator is linear, and if our light curve is given by the superposition of a sinusoidal signal and random noise, i.e. $x(t)=s(t)+n(t)$, the resulting Fourier transform will be the sum of the transforms of the two components, i.e. $\hat{X}=\hat{S}+\hat{N}$, where $\hat{S}\propto\delta(f-f_0)$ for a sinusoidal component with temporal frequency $f_0$. If the light curve is sufficiently long, an analysis in the frequency domain will identify the periodicity, even if a sinusoidal signal with a small amplitude will not manifest in the light curve due to a low signal-to-noise ratio. This is evident in Fig.~\ref{fig:lcps}, where we simulated a white noise-dominated time series, superposed to a signal with amplitude ten times smaller than the noise. While the inspection of the time series does not show any evidence for the presence of a periodicity, the power spectrum shows an evident peak at $f_0=10^{-3}$ Hz, corresponding to the simulated sinusoid frequency.

\subsection{Computing the significance above underlying white noise}
Depending on the power of the periodic component with respect to the underlying noise, the peak in the periodogram will be more or less prominent. Typically, the more periods are sampled in the light curve, the more prominent the periodic peak will be in the frequency domain, but quantifying its significance is not trivial. A common tool to estimate the significance of a peak above the underlying white noise is the so-called {\it Fisher's exact g-test} \citep{fisher29}. If one wants to analyze a peak at the frequency $f_0$, the test consists in computing the value $$g_0=\frac{P_\text{S}(f_0)}{\sum_{k=1}^NP_\text{S}(f_k)}$$ where $N$ is the number of points in our periodogram. The value $g_0$ corresponds to a p-value of

\begin{equation}
p_\text{val}=P(g>g_0)=\sum_{k=1}^b(-1)^{k-1}\frac{N!}{k!(N-k)!}(1-kg_0)^{N-1}
\label{eq:pvalgtest}
\end{equation}

\noindent where $b=\text{floor}(1/g_0)$, i.e. the largest integer smaller than $1/g_0$. Given the presence of factorial calculations, Eq.~\ref{eq:pvalgtest} is computationally impractical for very long light curves. We tested that our software is able to compute g-test p-values for light curves with up to $\sim5\times10^3$ data points. Computing p-values of longer time series hence requires other tests to be used, but it is beyond the scope of this paper.

\section{Simulations of colored noise light curves}
\label{sec:method}

\begin{figure}
\centering
\includegraphics[scale=0.47]{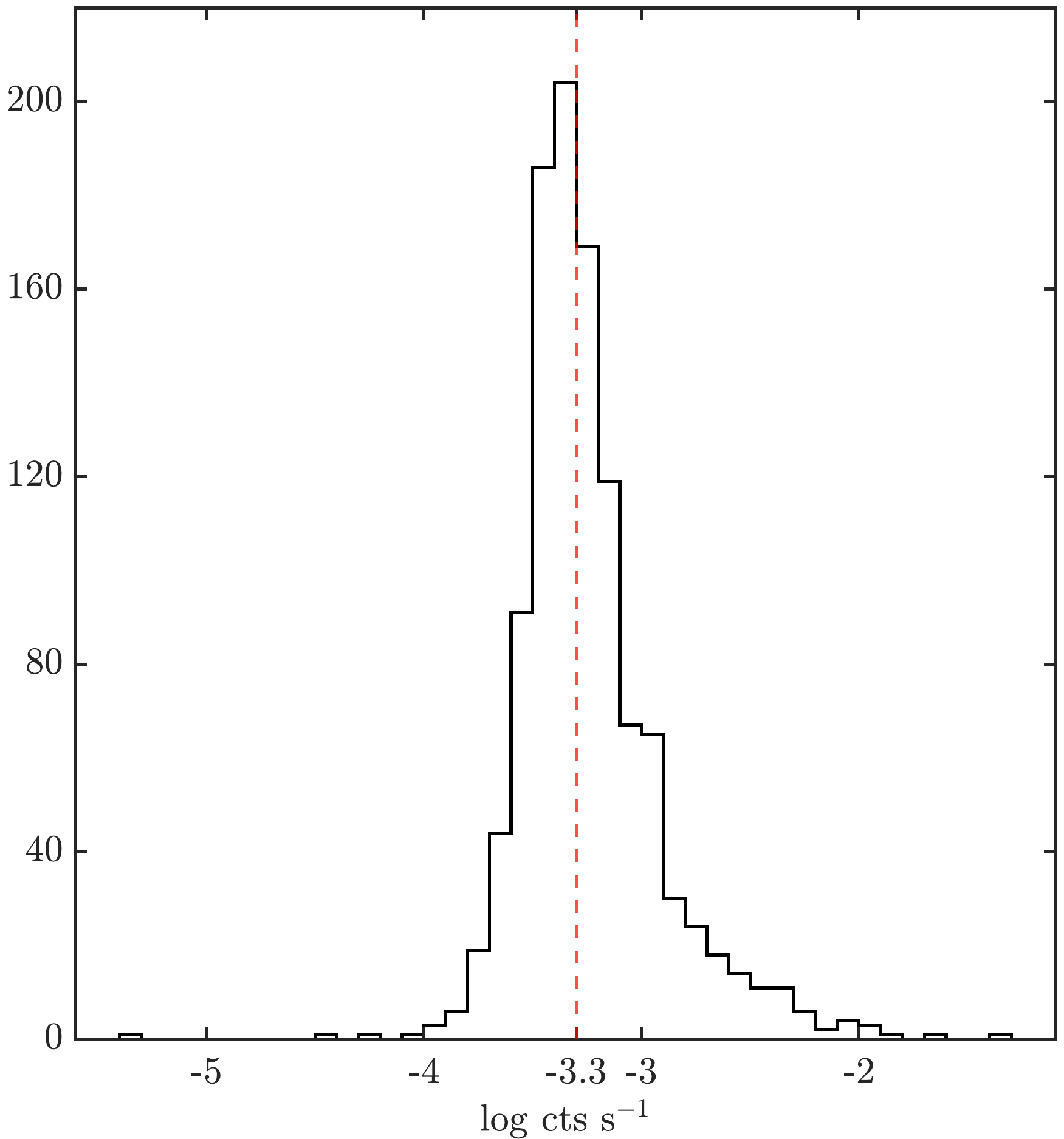}
\caption{Distribution of mean $14-195$ keV count rates of each {\it Swift}-BAT light curve of the whole 1103 AGN sample. The vertical red dashed line is the median value, corresponding to $\log(\text{CR}/\text{cts s}^{-1})\simeq-3.3$}
\label{fig:histrate}
\end{figure}

\begin{figure}
\centering
\includegraphics[scale=0.47]{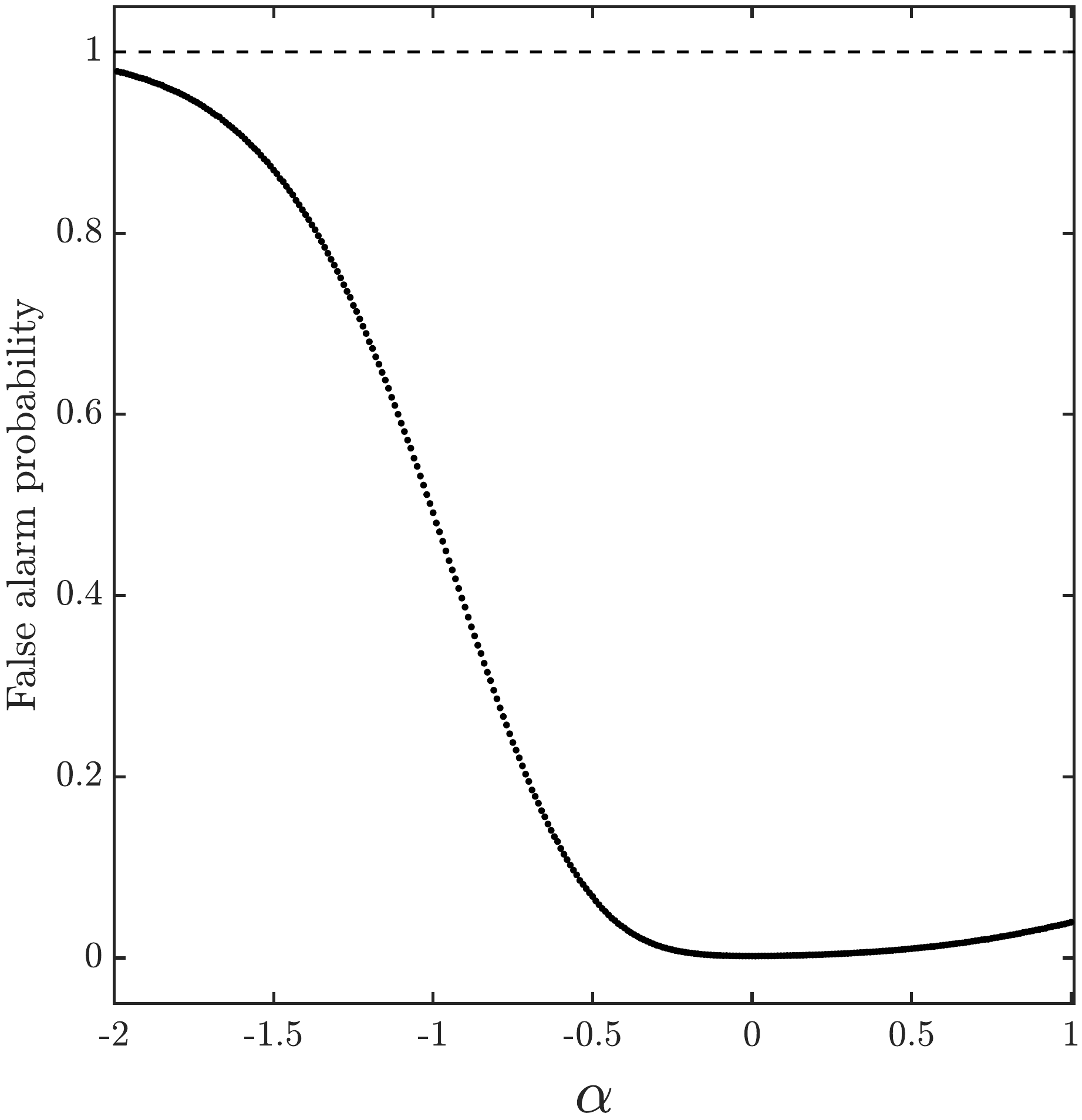}
\caption{False alarm probability as a function of the power spectral slope $\alpha$. For each value of $\alpha$ in the range $[-2\;1]$, with $\delta\alpha=0.01$, we simulate $10^6$ light curves, without injecting a periodic signal. For a given $\alpha$, the false alarm probability is the number of curves with peaks with p-values lower than the chosen, i.e. $p_\text{val}\leq p_\text{th}=0.0018$.}
\label{fig:fapalpha}
\end{figure}

The validity of the g-test is unfortunately limited to the cases in which the non-periodic noise is white. In fact, the power spectrum of the stochastic component of an AGN X-ray light curve is typically modelled as $P_\text{S}(f)\propto f^\alpha$, in absence of slope breaks in the sampled frequency range. In case of white noise PSD, the slope $\alpha$ will be approximately zero. However, it is well known that the underlying noise in X-ray light curves of AGN is not white, finding typical colored noise distributions, with slope values of $\alpha\sim-1.5$ in the frequency range $\log(f/\text{Hz})\gtrsim-6$ \citep[e.g.,][]{green93,lawrence93,allevato13,emmanoulopoulos16}. Ignoring the fact that the noise is colored can lead to either false positives \citep[e.g.,][]{vaughan16}, because the peak is tested against a flat noise instead of a $f^\alpha$ noise, or hide true positives, which may be below overwhelmingly large noise fluctuations at low frequency. In such cases the result of the g-test cannot be used as it is and we need a way to take the underlying slope of the PSD into account.\\
\indent In order to evaluate the validity of the g-test in a colored noise PSD, we run simulations to assess the false alarm rate caused by the colored noise. To date, the only X-ray survey that at least partially samples the temporal frequency of interest ($f\sim10^{-9}-5\times10^{-8}$ Hz, corresponding to orbital periods of $P\sim9-400$ months), matching the frequency range where PTAs are most sensitive \citep[e.g.,][]{lentati15,moore15,shannon15,mingarelli17,arzoumanian18}, is the {\it Swift}-BAT 105 Month Survey \citep{oh18}. The survey provides Crab-weighted light curves binned at one month for 1631 sources, detected at $>5\sigma$ confidence level \citep[see][for details]{baumgartner13}. It covers almost half the sky at the energy range $E=14-195$ keV and is therefore unlikely affected by neutral or ionized absorbers, soft excess or other spectral features not associated with the primary emission. The analysis of the real {\it Swift}-BAT data is presented in Section~\ref{sec:data}.\\ 
\indent  We base our simulations on mock light curves that have the same characteristics of the real data that we intend to analyze. We only consider the AGN in the catalog, whose average count rates are distributed accordingly to Fig.~\ref{fig:histrate}. We then select only the brightest half of the sample, rightwards of the vertical red dashed line that represents the median count rate value $\log ({\rm CR}/{\rm cts}\;{\rm s}^{-1})\simeq-3.3$. (see Sect.~\ref{sec:data}), counting a total of $553$ AGN.\\
\indent We simulate 105-months long light curves with Gaussianly-distributed count rates, i.e. white noise light curves. The mean of the Gaussian distribution is a typical value in the considered count rate range, while the standard deviation is the mean value of the photometric error. We verify that the results do not depend on the particular mean value we adopt in the range $-3.3\lesssim\log ({\rm CR}/{\rm cts}\;{\rm s}^{-1})\lesssim-1.5$, nor on the average value of the photometric error.\\
\indent For $N$ sources with white noise light curves, the probability of finding one false periodicity by chance is given by $p({\rm FA|\alpha=0})=1/N$, which is $p=0.0018$ for $N=553$. Therefore, the g-test significance for white noise power spectra will be considered acceptable if the p-value satisfies $p_{\rm val}\leq p_{\rm th}=0.0018$. To check this procedure, we simulate $10^6$ pure white noise light curves, i.e. without any injected periodicity, we perform g-tests on each one and we flag every source with $p_{\rm val}\leq p_{\rm th}$. The false alarm probability is given by the number of flagged light curves over the total number. As expected, we find $p({\rm FA}|\alpha=0)\simeq0.0018$. This is clearly not true for $\alpha\neq0$, therefore we need to simulate light curves for different values of $\alpha$ and thus calculate $p({\rm FA}|\alpha)$.\\
\indent In order to simulate colored noise light curves, we adopt the technique presented by \citet{kasdin95}, which we summarize in the following. We simulate a white noise light curve $x^{(w)}(t)$ as previously described, and its Fourier transform is $\hat{X}^{(w)}(f)\propto f^0$. It is possible to simulate a colored noise light curve, described by its spectral slope $\alpha$, by defining the filter $\hat{H}(f)=f^{\alpha/2}$. We can introduce the quantity $\hat{X}^{(c)}=\hat{X}^{(w)}(f)\times\hat{H}(f)\propto f^{\alpha/2}$, whose inverse Fourier transform is $x^{(c)}(t)$. The periodogram of such light curve is $P_\text{S}(f)=\left|\hat{X}^{(c)}(f)\right|^2\propto f^\alpha$, therefore by construction the simulated light curve $x^{(c)}(t)$ is made up of a stochastic colored noise process with spectral slope $\alpha$.\\
\indent These simulations can be used to quantify the expected number of false alarms in colored noise curves. We consider $\alpha$ in $\left[-2,\;1\right]$, spaced by $\delta\alpha=0.01$. For each value of $\alpha$ we simulate $10^6$ light curves following the above described procedure. Again, for each value of $\alpha$, we count the light curves with $p_{\rm val}\leq 0.0018$ and divide such number by $10^6$, hence tracing $p({\rm FA}|\alpha)$ in the selected interval, as shown in Fig.~\ref{fig:fapalpha}.


 


\section{Analysis of real {\it Swift}-BAT data}
\label{sec:data}

\subsection{Identification of possible periodic candidates}


\begin{figure}
\centering
\includegraphics[scale=0.45]{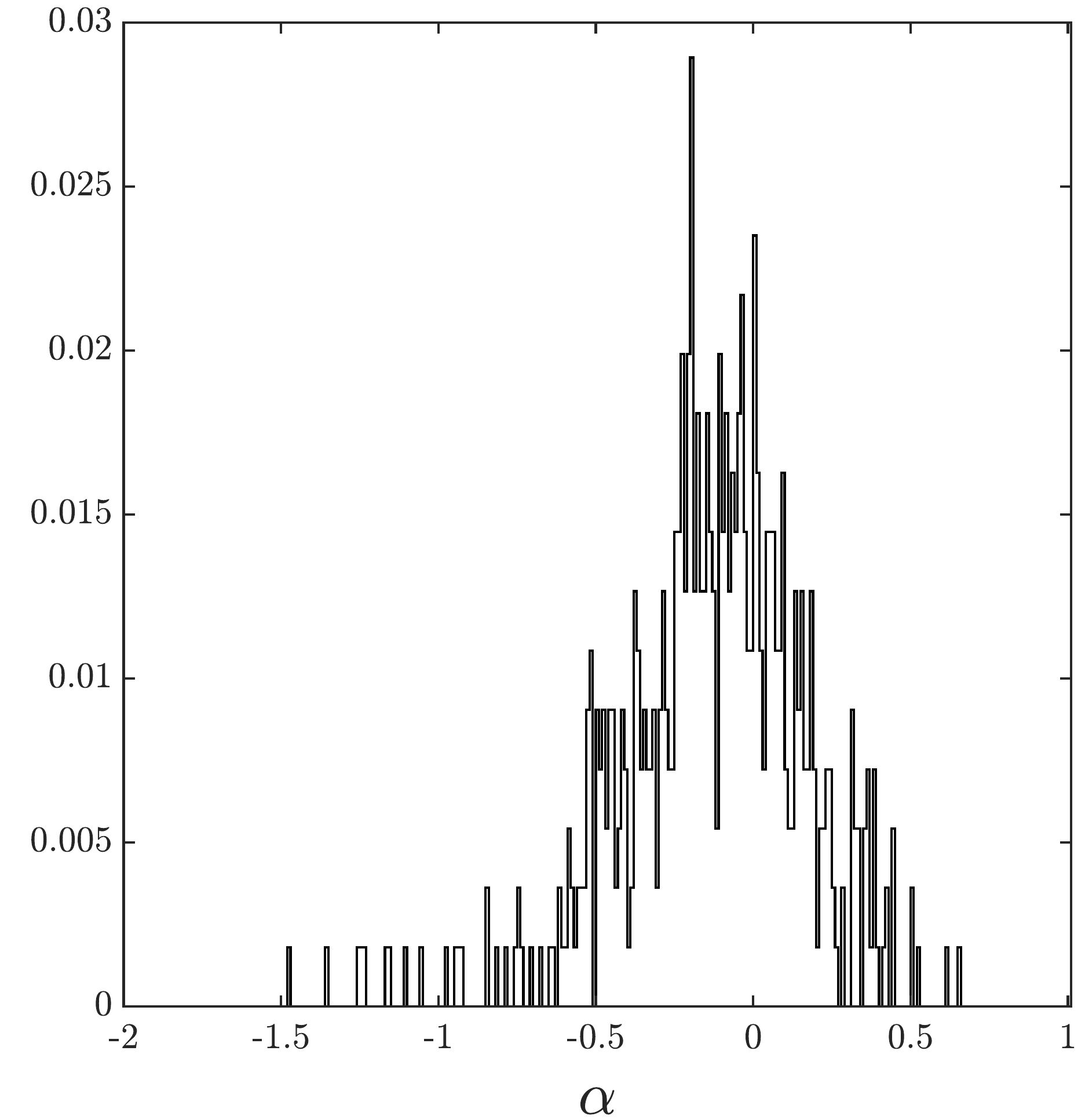}
\caption{Distribution of the PSD slope $\alpha$ for the $553$ AGN sources considered in this work. The distribution is normalized to unity in order to represent the probability distribution function of finding a light curve with slope $\alpha$.}
\label{fig:histalpha}
\end{figure}

\begin{figure}
\centering
\includegraphics[scale=0.245]{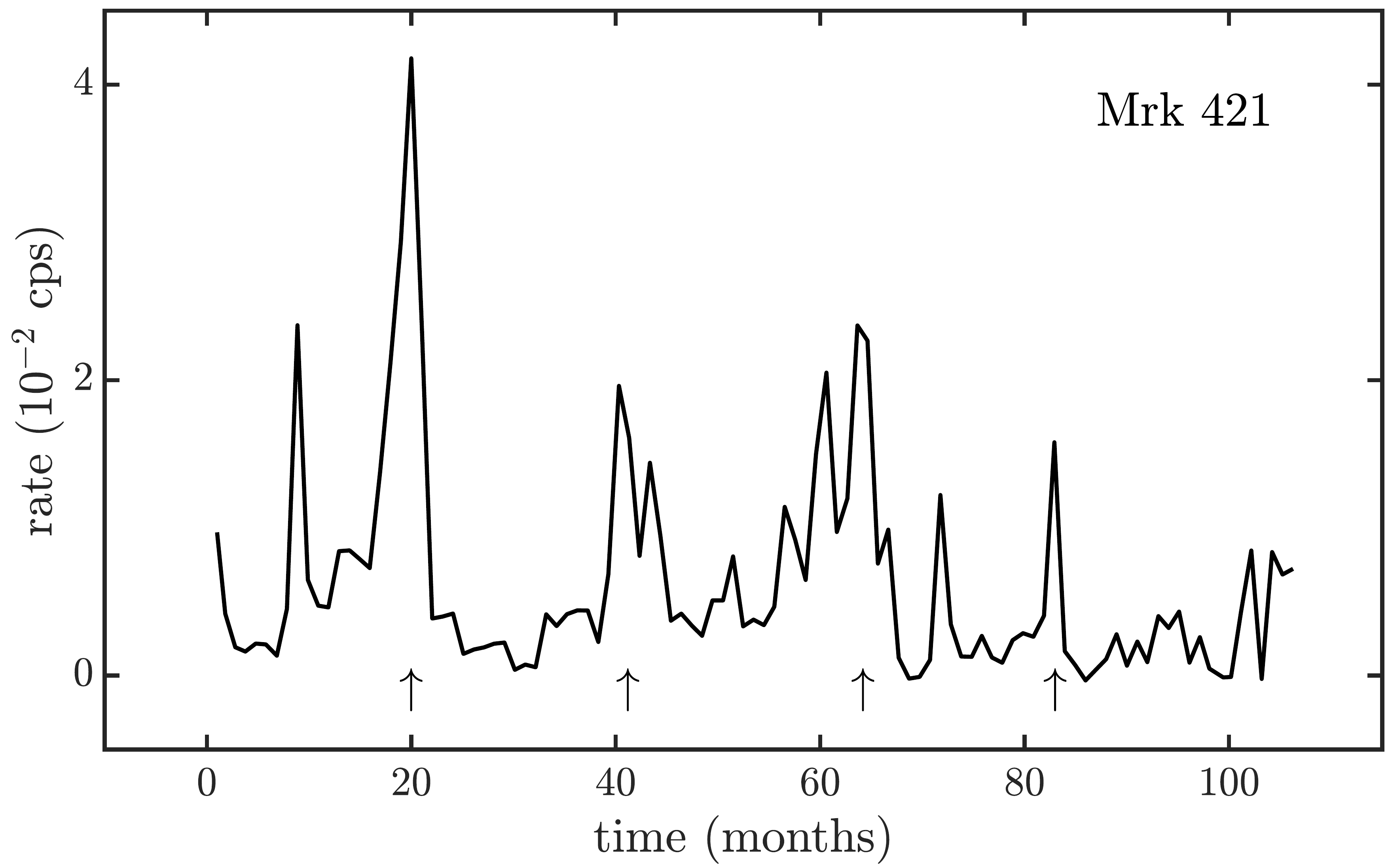}
\caption{Hard X-ray light curve of Mrk 421. The errors are omitted, since they are extremely small. Points are connected to highlight the flaring nature of the light curve variability. The arrows indicate four possible flares, happening at $\sim21$ months from each other, with different intensity.}
\label{fig:mrk421_lc}
\end{figure}

A first step to identify possible candidates is to perform g-tests on the sample of $553$ {\it Swift}-BAT sources presented above, using the following multi-step approach: i) we do not make any assumption on the nature of the stochastic noise of the light curves, i.e. we assume that they are all characterized by white noise. ii) We identify the highest PSD peak, and we test such peak against the rest of the PSD, assumed as noise. We only consider peaks at periods $9\lesssim P_0\lesssim35$ months. The lower limit is set by the PTA sensitivity curve \citep[e.g.,][]{lentati15}, while the upper limit is set by the requirement that we observe at least three phases in a 105-month light curve, i.e. $105/3=35$ months. iii) We select all those sources whose p-value satisfy the condition $p_\text{val}\leq1/553=0.0018$, which is the threshold corresponding to the probability of finding one false positive by chance in the total sample analyzed, without setting any prior. iv) Once the candidates are identified, we investigate  whether we are dominated by colored noise fluctuations or not, and then estimate the probability that the periodic component is real.\\
\indent Only two sources survive the first screening. The first one is the blazar Mrk 421, which has a PSD peak at $f_0=18\pm2$ nHz, corresponding to a period of $P_0=21\pm2$ months, with $p_\text{val}=5\times10^{-4}$. The other selected AGN is the Seyfert 1.5 galaxy Mrk 915, with a PSD peak at $f_0=11\pm2$ nHz, corresponding to the period $P_0=35^{+7}_{-5}$ months, with $p_\text{val}=8\times10^{-4}$, a factor about $3.6$ and $2.3$ below the threshold $p_{\rm val}=0.0018$, respectively. We note that the uncertainties are given by the frequency resolution of the PSD at the peak frequency $f_0$.\\
\indent The initial screening does not take into account the intrinsic color of the stochastic noise, i.e. the slope of the PSD. Assuming a simple power-law model with no breaks in the sampled frequency range for the PSD, i.e. $P_S(f)\propto f^\alpha$, the slope can be computed with linear fits between the logarithmic values of $\log P_S$ and $\log f$ \citep[e.g.,][]{papadakis93}. We compute $\alpha$ for each source, with its normalized distribution shown in Fig.~\ref{fig:histalpha}. We note that the bulk of our sample has PSD slopes $-0.5\lesssim\alpha\lesssim0.5$, with a tail of very few sources characterized by $\alpha\lesssim-0.5$. Even though the typical slope at higher frequencies is $\alpha\sim-1.5$ \citep[e.g.,][]{allevato13}, this result is not surprising, since the X-ray PSD of AGN is expected to break to lower slopes at frequencies $f\lesssim10^{-6}$ Hz \citep{shimizu13}, with PSD breaks observed in a few sources so far \citep{markowitz03,mchardy07,macleod10}. For this sample, a break timescale smaller than the Nyquist period, i.e. $T_\text{br}\lesssim2$ months (or $f_\text{br}\gtrsim2\times10^{-7}$ Hz) for the $98\%$ of the sources has been estimated by \citet{liu20}, which means that the PSD break happens at much shorter timescales than those sampled by {Swift}-BAT, explaining the distribution of $\alpha$ shown in Fig~\ref{fig:histalpha}.\\
\indent The slopes of the PSDs for the two g-test selected candidates, Mrk 421 and Mrk 915, are $\alpha=-1.2\pm0.2$ and $\alpha=-0.4\pm0.3$, respectively. This means that Mrk 421 is one of those outliers that mantains a red noise behavior even at extremely low frequencies, while Mrk 915 is characterized by almost white noise. We estimate the number of expected false alarms as the product of the false alarm probability for a given $\alpha$, $p(\text{FAP}|\alpha)$ (see Fig.~\ref{fig:fapalpha}), and the probability of $\alpha$, $p(\alpha)$ (see Fig.~\ref{fig:histalpha}), multiplied by the number of sources $N=553$, i.e. $N_\text{exp}=N p(\text{FAP}|\alpha) p(\alpha)$. Concerning Mrk 915, for $\alpha=-0.4$ we re-simulate $p(\text{FAP}|\alpha=-0.4)$, using $p_\text{val}=8\times10^{-4}$ as threshold, obtaining $N_\text{exp}\simeq0.1$ expected false alarms. As for Mrk 421 ($\alpha=-1.2$, $p_\text{val}=5\times10^{-4}$), we obtain $N_\text{exp}\simeq0.7$. This means that, assuming Poissonian distributions with expected value $N_\text{exp}$, the probability that Mrk 915 is a false positive is $\sim6\%$, while for Mrk 421 such probability is $\sim32\%$.\\
\indent The high chance of Mrk 421 of being a non-periodic component is supported by the fact that the source is well-known to recurringly flare in all bands \citep[e.g.,][]{aleksic15}, including the X-rays \citep[e.g.,][]{stroh13,hervet19}. As shown in Fig.~\ref{fig:mrk421_lc}, the source has numerous flares of various intensities at $\sim20$, $\sim41$, $\sim63$ and $\sim83$ months from the beginning of the survey. It is therefore straightforward to conclude that the PSD is indeed identifying a recurring variability pattern with a typical timescale of $P_0\sim21$ months, but it is clearly not due to a proper sinusoidal behavior of the light curve. Moreover, it is well-known that, for blazars like Mrk 421, the dominant X-ray emission mechanism is related to the presence of a jet rather than accretion \citep[e.g.,][]{fossati98,ghisellini17}.\\
\indent We conclude that Mrk 421 is not a binary candidate, according to our selection criteria, while Mrk 915 deserves a closer look at its light curve.

\subsection{Light curve analysis of Mrk 915}
\label{sec:mrk915}

\begin{figure}
\centering
\includegraphics[scale=0.45]{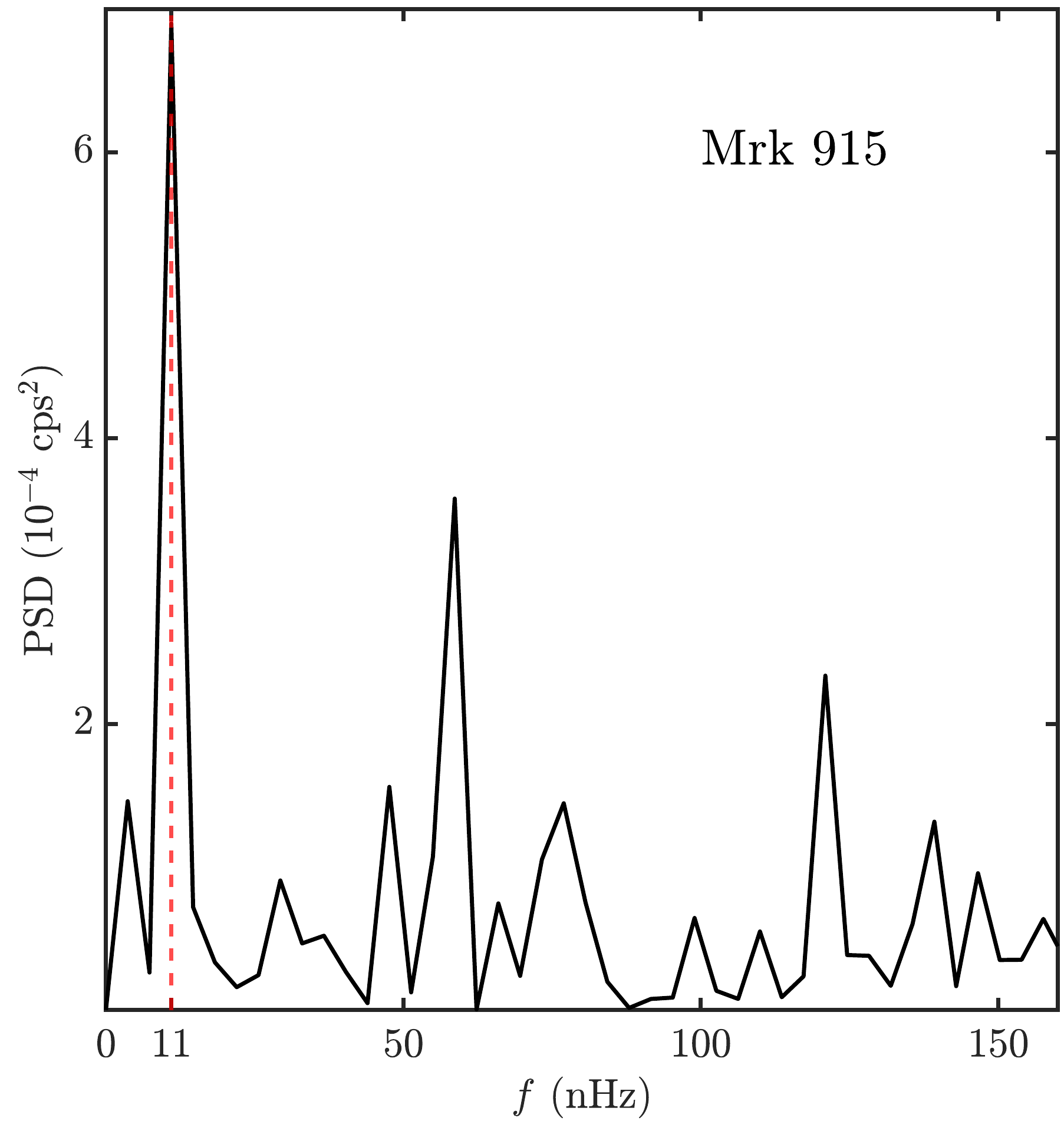}
\caption{Power spectral density of Mrk 915. The red vertical dashed line represents the peak at $f_0=11\pm2$ nHz (or $P_0=35^{+7}_{-5}$ months).}
\label{fig:mrk915_psd}
\end{figure}

\begin{figure}
\centering
\includegraphics[scale=0.245]{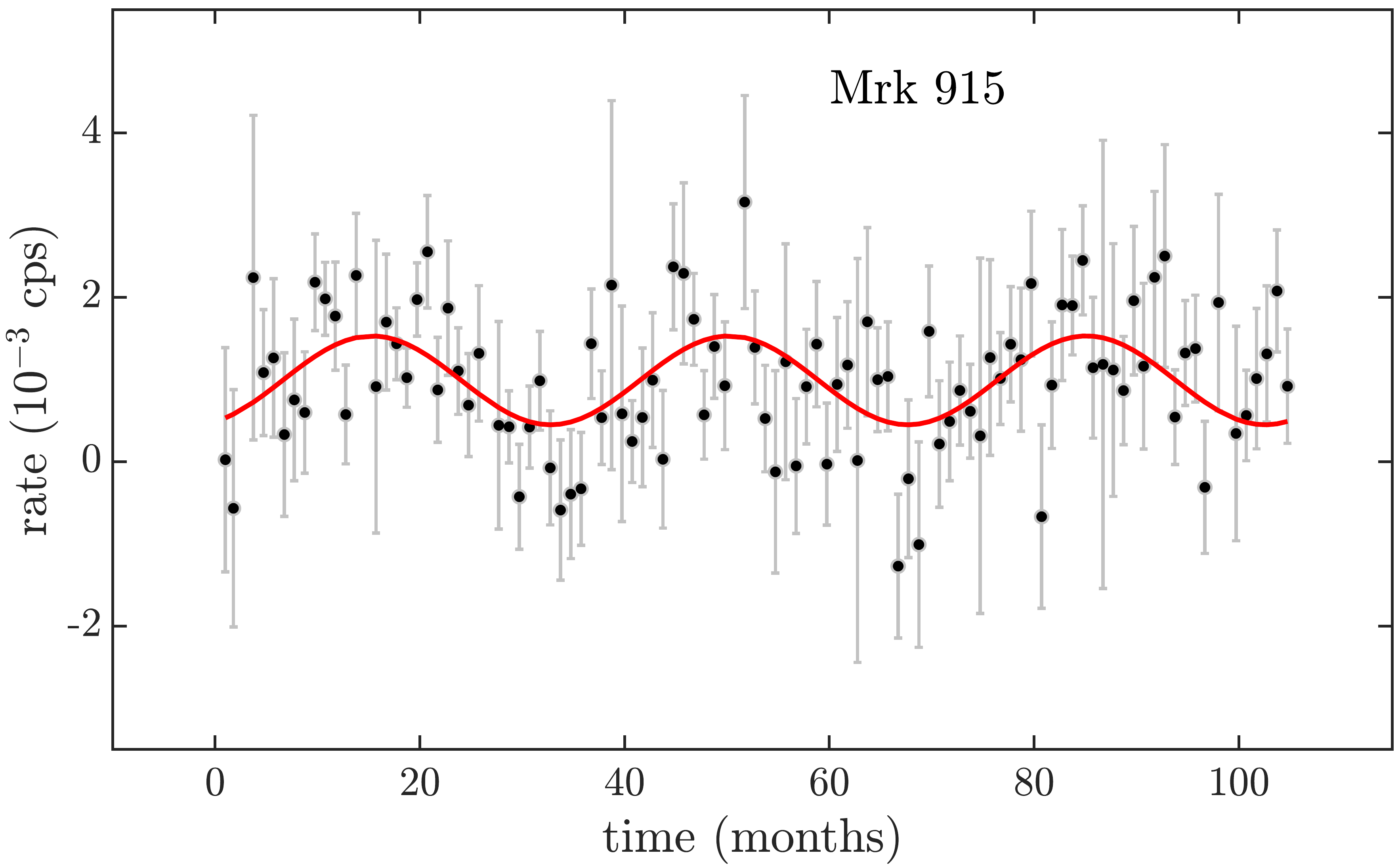}
\caption{Hard X-ray light curve of Mrk 915. The red line represents the best-fit sinusoidal curve.}
\label{fig:mrk915_lc}
\end{figure}

\begin{figure}
\centering
\includegraphics[scale=0.45]{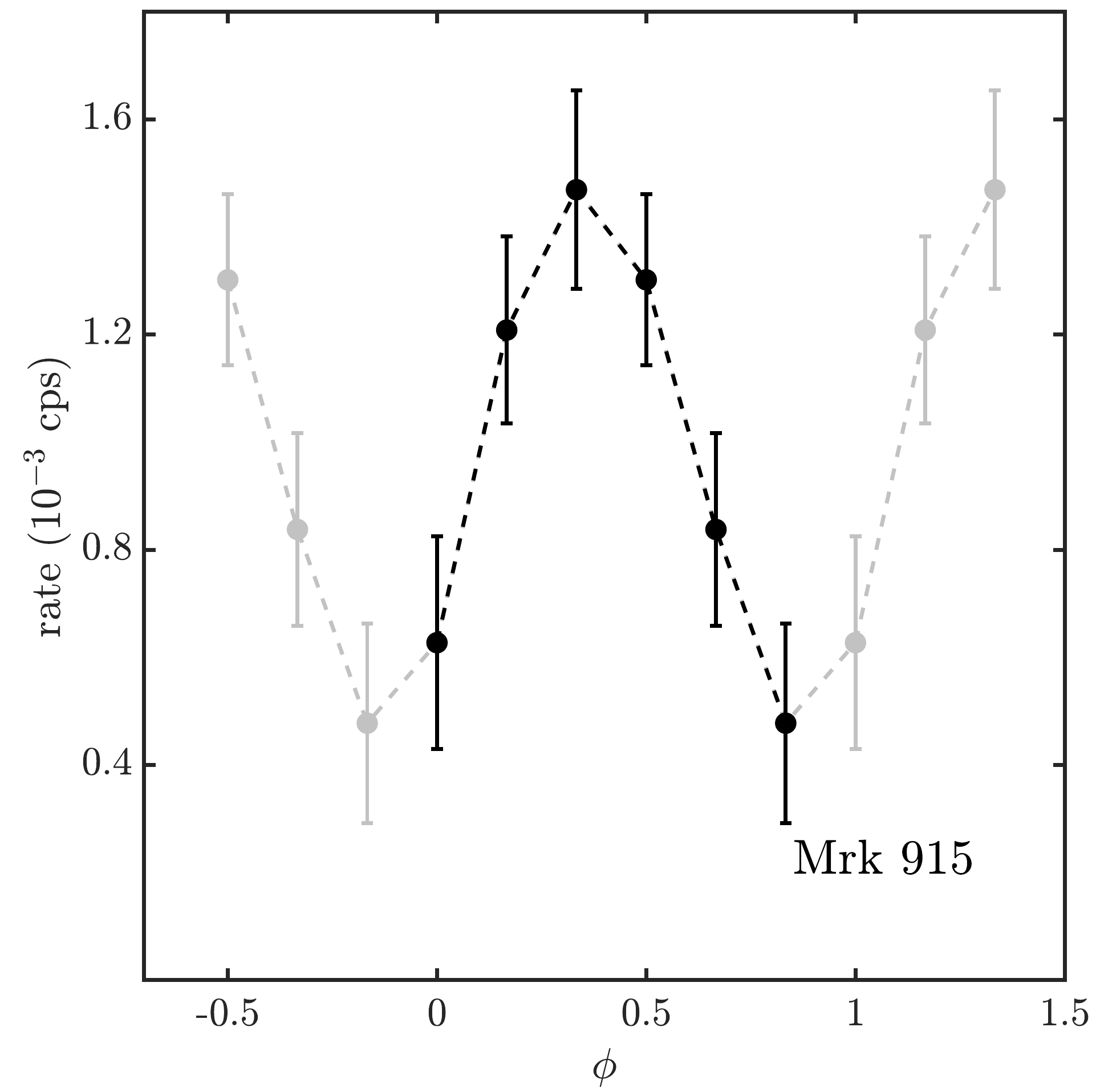}
\caption{Phase-folding of the Mrk 915 hard X-ray light curve. The folding period was divided into six phase bins, each one combining $\sim16$ light curve data points. The grey points are mirrored from the main six phase bins, in order to highlight the periodical trend.}
\label{fig:mrk915_ef}
\end{figure}

As reported in the previous section, the variability pattern of Mrk 915 possibly includes a periodic component with frequency $f_0=11\pm2$ nHz, corresponding to a period of $P_0=35^{+7}_{-5}$ months (see Fig.~\ref{fig:mrk915_psd}). However, the g-test does not allow us to assess an absolute statistical significance, since it is tested near the boundary condition \citep[see][for details]{protassov02}, so it is only used to screen potential interesting candidates in our work. Moreover, the g-test does not take the photometric errors into account and, therefore, in case the errors are large, i.e. their excess variance is small or negative, the periodicity might be apparent and must be tested with other methods. In the specific case of Mrk 915, the excess variance is negative, which implies that the observed variability could be apparent, and hence the PSD analysis must be handled with care and the putative periodicity should be investigated further.\\
\indent In order to investigate this and to take the errors into account we first attempt a simple weighted least-squared sinusoidal fit of the light curve.\footnote{We remove four points associated with unusually low exposure times, and thus anomalous large errors, during that specific month.} The fitted function is given by $y(t)=A_0+A_1\sin(2\pi P_0^{-1}(t-t_0))$, where $A_0$, $A_1$, $t_0$, and $P_0$ are all free parameters. We obtain the following best-fit values, with errors given at $90\%$ confidence level: $$A_0=(10\pm2)\times10^{-4}\;{\rm cps}$$ $$A_1=(5\pm2)\times10^{-4}\; {\rm cps}$$ $$t_0=53\pm3\;{\rm months}$$ $$P_0=35\pm2\;{\rm months}.$$ The goodness of fit is given by $\chi^2/{\rm dof}=82/96$. In Fig.~\ref{fig:mrk915_lc} we show the {\it Swift}-BAT light curve, overlapped by its best-fit sinusoidal curve. We note that the best-fit value of $P_0$ is consistent with the corresponding value of the PSD peak that is found with the g-test sample screening.\\
\indent We use this latter result to perform a phase-folding procedure on the light curve of Mrk 915 \citep[e.g.,][]{leahy83}. The epoch folding search is typically used to determine the period and the amplitude of the sinusoidal trend \citep[see e.g.,][]{leahy87}. However, given the short effective exposure time and the large period ($P_0\simeq35$ months), it is not possible to oversample the Fourier resolution ($\delta f=1 / \delta t_{exp}$). Therefore, we use $P_0$ as known parameter to fold the dataset with $n=6$ phase bins and test the null hypothesis. The number of bins is chosen as a trade-off between a high signal-to-noise ratio for each bin, and the fact that significance tests, such as the $\chi^2$ test, are more reliable with more data points available.\\
\indent We first compute the relative phase of each monthly observation as $$\phi = \frac{t}{P_0} - \text{int}\left(\frac{t}{P_0}\right),$$ where $t$ is the time relative to the points of the light curve shown in Fig.\ref{fig:mrk915_lc}, and $\text{int}(a)$ means the integer part of $a$. We then combine the count rates corresponding to all points belonging to the same phase bin with the relative 1$\sigma$ uncertainty. Given a set of $N$ observations with count rate $x_i\pm\delta x_i$ and exposure $\delta t_i$, the count rates are combined as their weighted mean $$x_{\rm bin}=\frac{\sum_{i=1}^N x_i \left( \frac{1}{\delta x_i}\right )^2}{\sum_{i=1}^N\left( \frac{1}{\delta x_i}\right )^2},$$ while the errors are given by $$\delta x_{\rm bin}=\frac{1}{\sqrt{\sum_{i=1}^N \left( \frac{1}{\delta x_i}\right )^2}}.$$
This procedure is repeated for all the six phase bins we divide the period into. The folded light curve is shown in Fig.\ref{fig:mrk915_ef}.\\
\indent To confirm the presence of a periodic trend, we test the null hypothesis with a $\chi^2$ test on the folded light curve, obtaining $\chi^2\simeq24$, corresponding to a p-value of $2.3 \times 10^{-4}$ for 5 degrees of freedom (i.e. 6 phase bins minus 1), which means that we are able to reject the null hypothesis with a confidence level of $3.7\sigma$.

\section{Light curve analysis of MCG+11-11-032}
\label{sec:mcg}

\begin{figure}
\centering
\includegraphics[scale=0.45]{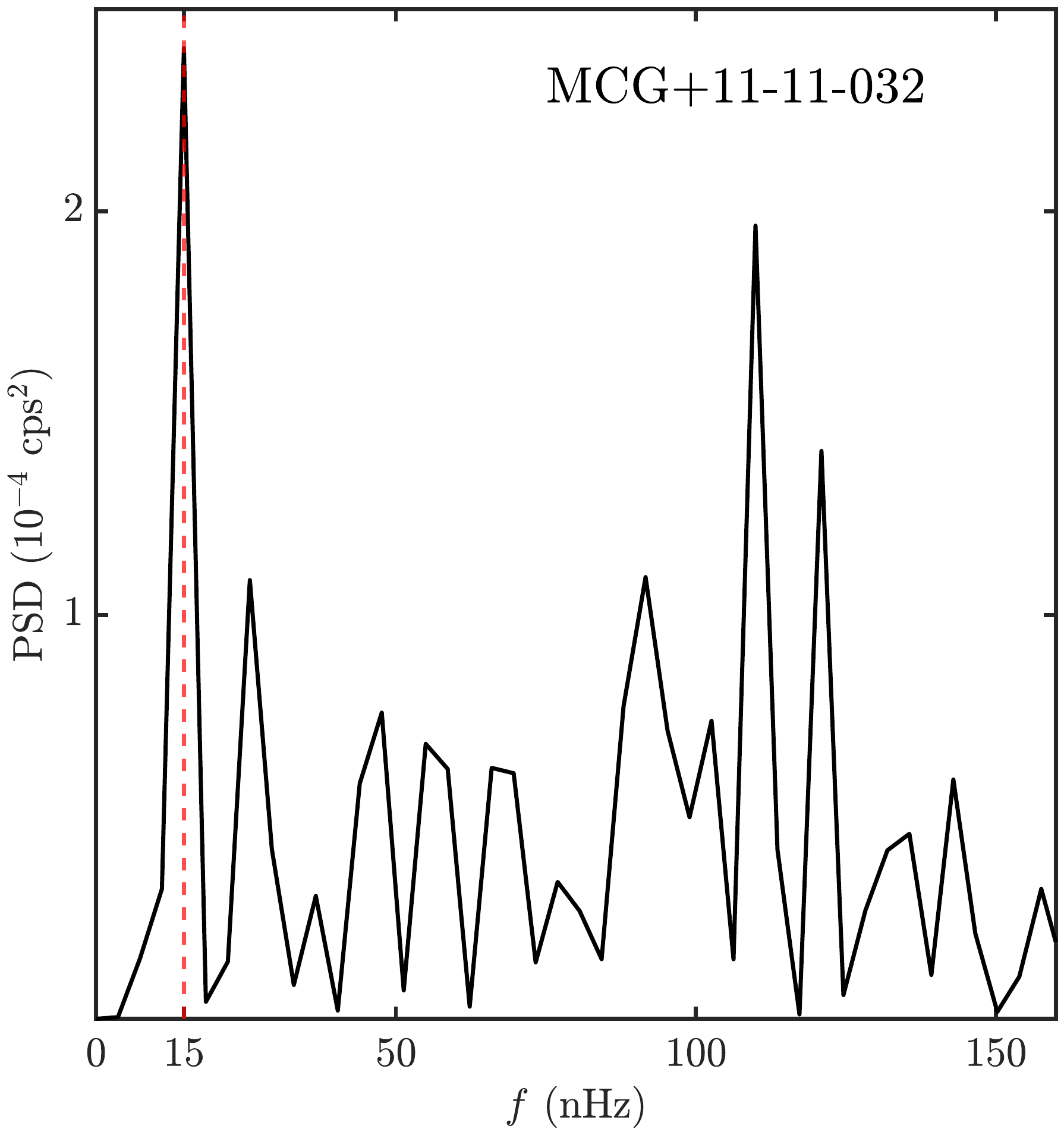}
\caption{Power spectral density of MCG+11-11-032. The red vertical dashed line represents the peak at $f_0=15\pm2$ nHz (or $P_0=26^{+4}_{-3}$ months).}
\label{fig:mcgpsd}
\end{figure}

\begin{figure}
\centering
\includegraphics[scale=0.45]{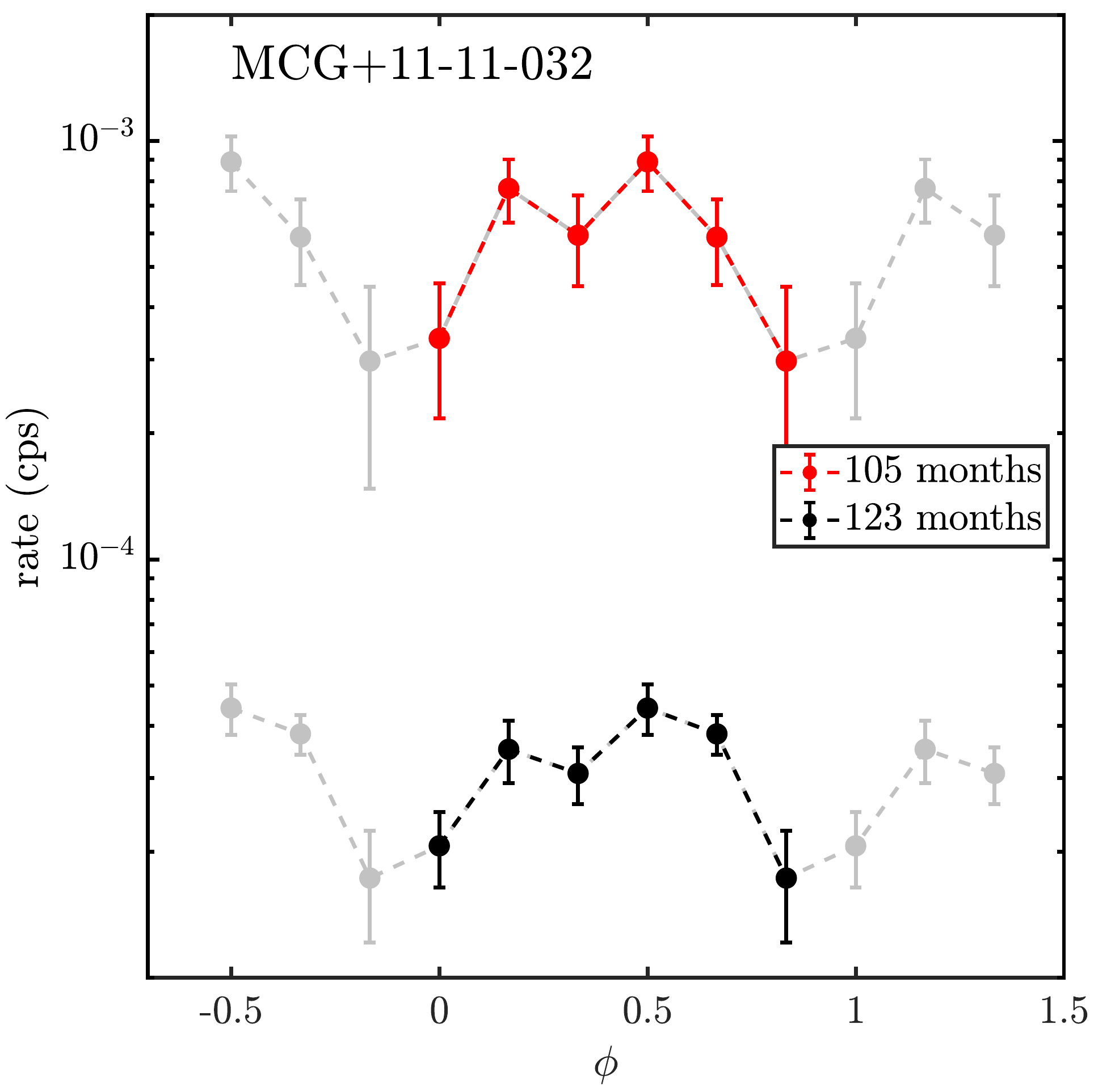}
\caption{Phase-folding of the MCG+11-11-032 hard X-ray light curve. The folding period was divided into six phase bins. The grey points are copies of the main six bins, in order to highlight the periodic behavior. The red points are the result of the epoch folding of the 105-Month {\it Swift}-BAT survey light curve, while the black points represent the folding of the {\it Swift}-BAT 123 months long light curve, in the $E=15-150$ keV energy band. The two curves have different count rates, because of different normalizations \citep[see][for further details]{baumgartner13,segreto10}.}
\label{fig:mcgef}
\end{figure}

As shown in Sect.~\ref{sec:data}, our method only selects one SMBHB candidate from the AGN sample. It must be stressed, though, that the method is very conservative, and aims at selecting only the strongest candidates. Indeed, the only previously X-ray-selected SMBHB candidate \citep{severgnini18}, the Seyfert 2 galaxy MCG+11-11-032, is included in our {\it Swift}-BAT sample, but was excluded with the g-test selection criterium. It must be noted that the p-value threshold is low because we do not consider in this work periodicities with significance below $1/N_\text{sources}$. This does not mean that  a periodic trend, which may be hidden by the  noise, or hidden by large measurement errors, is not present. As shown in Fig.~\ref{fig:mcgpsd}, the highest value of the PSD is found at $f_0=15\pm2$ nHz, which corresponds to $P_0=26^{+4}_{-3}$ months, with $p_{\rm val}>0.0018$, and hence not significant. The PSD slope is given by $\alpha=0.4\pm0.3$ and is therefore consistent with white noise at $90\%$ confidence level. However, this Seyfert 2 galaxy was identified as a possible SMBHB candidate by means of the presence of two Fe K$\alpha$ emission lines at energies $E=6.16\pm0.08$ keV and $E=6.56\pm0.15$ keV, at $4\sigma$ and $2\sigma$ confidence level, respectively. Such lines, if emitted by two separate sources in the center of the AGN, would imply that they are red- and blue-shifted Fe K$\alpha$ emission lines by two minidisks, orbiting each other with a period of $P_0\sim25$ months.\\ 
\indent The possible presence of a double Fe K$\alpha$ emission line justifies an exception on the PSD screening and to investigate this source further. Therefore, we analyze the light curve of MCG+11-11-032, using the same procedure as Mrk 915. We perform a weighted sinusoidal fit, testing again the function $y(t)=A_0+A_1\sin(2\pi P_0^{-1}(t-t_0))$. The best-fit values, with errors given at $90\%$ confidence level, are $$A_0=(5.8\pm0.9)\times10^{-4}\;{\rm cps}$$ $$A_1=(2\pm1)\times10^{-4}\;{\rm cps}$$ $$t_0=63\pm3\;{\rm months}$$ $$P_0=26.3\pm0.6\;{\rm months,}$$
\noindent with a goodness of fit of $\chi^2/{\rm dof}=104/101$. The period derived is in agreement with the estimate given by \cite{severgnini18}. Furthermore, also in this case the period found with the sinusoidal fit is similar to the period corresponding to the maximum value of the PSD (Fig.~\ref{fig:mcgpsd}).\\
\indent To further investigate MCG+11-11-032, we use the procedure adopted in Sect.~\ref{sec:mrk915} to perform the epoch folding of the light curve of MCG+11-11-032, adopting six phase bins (red points in Fig.~\ref{fig:mcgef}).  Also in this case we test the null hypothesis using a $\chi^2$ test, obtaining $\chi^2\simeq15$, which corresponds to a p-value of $p_\text{val}=0.0097$ for 5 degrees of freedom. This means that the null hypothesis is rejected at $2.6\sigma$ confidence level.\\
\indent Assuming that the double Fe K$\alpha$ spectral feature and the sinusoidal trend are two different experiments of the same physical scenario, i.e. the presence of a supermassive black hole binary system, we use the meta-analysis technique called {\it Fisher's combined probability test} \citep{fisher48}. Given $k$ independent experiments, each resulting in its own p-value $p_{\text{val},i}$, the test consists in computing the following statistic

\begin{equation}
X_{2k}^{2}=-2\sum_{i=1}^{k}\ln(p_{\text{val},i}),
\label{eq:fishermeta}
\end{equation}

\noindent where $X_{2k}^{2}$ is distributed as a $\chi^2$ with 2$k$ degrees of freedom, and can be therefore tested as such. Since in our case $k=2$, we test against a $\chi^2$ distribution with $4$ degrees of freedom. In \citet{severgnini18} a significance of $2\sigma$ ($p_\text{val}=0.05$) was reported for the least significant emission line. Using Eq.~\ref{eq:fishermeta} to combine such p-value with the one obtained with the epoch folding we obtain $X^2_4=15.3$, corresponding to a global combined p-value of $p_\text{val}=0.0042$ for $4$ degrees of freedom. This means that, under the assumption that the spectral emission lines and the periodicity are both caused by the presence of a binary system, we are able to reject the null hypothesis of non-binarity with a $2.9\sigma$ confidence level.\\
\indent As shown in Fig.~\ref{fig:mcgef}, the M-shape of the epoch folding plot suggests the possible presence of two harmonics. We add a second harmonic component to the fitting model, i.e. we fit the function $y(t)=A_0+A_1\sin(2\pi P_0^{-1}(t-t_0))+A_2\sin(4\pi P_0^{-1}(t-t_0))$. We find a goodness of fit of $\chi^2/\text{dof}=101/99$, with an improvement given by $\Delta\chi^2/\Delta\text{dof}=3/2$, meaning that the addition of the second harmonic is not significantly required by the data. However, we note that the best-fit period, $P_0=25.9\pm1.5$ months, is consistent with the one found fitting only one sinusoidal component. A longer light curve, possibly with a lower noise level, would be required to confirm the presence of a possible second harmonic component.\\
\indent For this source, we also consider the 123-month $15-150$ keV Palermo {\it Swift}-BAT light curve \citep{segreto10}, available only for this AGN, which was already presented in \cite{severgnini18} (Palermo {\it Swift}-BAT team, private communication). Assuming the  derived period of $\sim26$ months, this longer light curve thus samples almost five periods and could in principle return a higher significance for rejecting the null hypothesis. We bin the light curve at 5 months of observations per data point, to maximize the signal-to-noise ratio. The curve is well fitted by a single sinusoid $y(t)=A_0+A_1\sin(2\pi P_0^{-1} (t-t_0))$, with a goodness of fit of $\chi^2/{\rm dof}=17/19$, and a best-fit period $P_0=26.0\pm0.6$ months, consistent with the best-fit obtained with the $105$-month light curve. We use such period to undertake the epoch folding technique, shown in Fig.~\ref{fig:mcgef} (black points). We test the null hypothesis with a $\chi^2$ test, obtaining $\chi^2\simeq21$, which corresponds to $p_{\rm val}=7.5\times10^{-4}$, implying that for this longer light curve the null hypothesis can be rejected at the $3.4\sigma$ confidence level. Considering also the result in \cite{severgnini18} in the combined probability test (Eq.~\ref{eq:fishermeta}), we obtain $p_{\rm val}=4.2\times10^{-4},$ which means that the null hypothesis can be rejected with a $3.5\sigma$ confidence level. The black points in Fig.~\ref{fig:mcgef} suggest also in this case the presence of a  second harmonic. However, adding a second harmonic component in the light curve fit still produces a consistent period ($P_0=26.1\pm0.8$ months), but the goodness of fit is $\chi^2/{\rm dof}=12.6/17$, hence the second harmonic is required by the F-test at only $93\%$ confidence level.

\section{Discussion and conclusions}
\label{sec:summary}

\begin{table*}
\centering
\begin{tabular}{cccccccc}
\hline
Source Name & Type  & Identified as candidate & Period (months) & $\chi^2$ p-value & $d_L$ (Mpc) & $a$ ($10^{-3}$ pc) & $\Delta v/c$\\
\hline
Mrk 915 & Seyfert 1.5 & This paper & $35\pm2$ & $2.3\times10^{-4}$ & $105$ & $5\pm1$ & $0.034\pm0.007$\\
MCG+11-11-032 & Seyfert 2 & \cite{severgnini18} & $26.3\pm0.6$ & $9.7\times10^{-3}$ & $160$ & $6.5\pm1.5$ & $0.06\pm0.02$\\
\hline
\end{tabular}
\caption{Main properties of the two sources analyzed in this paper. We report the name of the source, the AGN type, the reference where it was first reported as candidate SMBHB, the period obtained with a sinusoidal fit in units of months, the epoch-folding p-value obtained with a $\chi^2$ test of the null hypothesis, the luminosity distance of the AGN, the separation between the two black holes in the assumption that the observed periodicity is due to a binary system, and their relative velocity.}
\label{tab:recap}
\end{table*}

We have presented a method to select possible SMBHB candidates using the Fisher's exact g-test, finding a candidate in the Seyfert 1.5 galaxy Mrk 915 with period given at the maximum value of the PSD of $P_0=35^{+7}_{-5}$ months. This source is not affected by colored noise, since the slope of its PSD is given by $\alpha=-0.4\pm0.3$, implying almost white noise, and there is only a $6\%$ probability of the periodicity to be a false positive. As a follow-up of this selection, we have performed a sinusoidal fit on the {\it Swift}-BAT hard X-ray light curve spanning 105 months, and we found a period of $P_0=35\pm2$ months, consistent with the period corresponding to the maximum PSD value, with a goodness of fit of $\chi^2/{\rm dof}=82/96$. Furthermore, we tested such period with an epoch folding procedure, finding that the null hypothesis can be rejected with a $3.7\sigma$ confidence level. Assuming a Keplerian orbit, the distance $a$ between the two X-ray-emitting regions that embed the two SMBHs can be inferred by using the third Kepler's law, i.e. $a=\left[GMP_0^2/4\pi^2 \right]^{1/3}$, where $M$ is the total mass of the two black holes, $P_0$ is the measured period, and $G$ the gravitational constant. We adopt the black hole mass estimate $M=(1.1\pm0.4)\times10^8\;M_\odot$, computed using single-epoch H$\beta$ measurements \citep{bennert06}. We derive a distance between the two hard X-ray-emitting regions of $a=(5\pm1)\times10^{-3}$ pc. Assuming a circular orbit, the relative velocity between the putative black holes is $\Delta v=2\pi aP_0^{-1}=(3.4\pm0.7)\times10^{-2}\;c$. Previous X-ray spectroscopy analyses of the source \citep[e.g.,][]{severgnini15,ballo17} have not unveiled any double Fe K$\alpha$ feature. In particular, XMM-Newton, with its energy resolution of $\sim0.15$ keV at the line energy, detects a single Fe K$\alpha$ line at $E=6.42\pm0.02$ keV with no double peaks \citep{ballo17}. We can therefore conclude that, if the binary system is able to produce a double peak, the energy separation between the two peaks would be $\Delta E\lesssim0.15$ keV. Given the orbital velocity $\Delta v$ obtained from the measured period, the expected Fe K$\alpha$ energy shift is given by $\Delta E\text{ (keV)}=E_{{\rm Fe K}\alpha}\Delta v/c\simeq0.2 \sin\iota$, where $E_{{\rm Fe K}\alpha}\simeq6.4$ keV and $\iota$ is the inclination angle. Since XMM-Newton only observes one emission line, if two peaks are indeed present in the X-ray spectrum, with $\Delta E\lesssim0.15$ keV, this would mean that the inclination angle is $\iota\lesssim\pi/4$. Alternatively, in case the source is indeed able to produce double Fe K$\alpha$ lines with energy separation above $\Delta E\sim0.15$ keV, the two black holes could have been observed in a state in which they are aligned and hence the radial component of the velocity could be negligible. However, this is highly unlikely since the source has been observed more than once by {\it Swift}-XRT, Suzaku and XMM-Newton at epochs that are not spaced by half period, and in none  of these observations a possible double Fe K$\alpha$ line was detected, which excludes the aligned black hole hypothesis.\\
\indent Given that MCG+11-11-032 was reported as a possible binary SMBH by \cite{severgnini18}, we also analyzed the light curve of this Seyfert 2 galaxy, finding a best-fit period of $P_0=26.3\pm0.6$ months, with a goodness of fit of $\chi^2/{\rm dof}=104/101$. By visual inspection, \cite{severgnini18} gave a rough estimate of the period of $P_0\sim25$ months, which is likely consistent with our result. From the SMBH mass of \cite{lamperti17}, $\log(M/M_\odot)=8.7\pm0.3$, derived from the velocity dispersion of the CO molecules, we infer that the separation of the two black holes is $a=\left[GMP_0^2/4\pi^2 \right]^{1/3}=(6.5\pm1.5)\times10^{-3}$ pc, from which we can quickly compute $\Delta v=2\pi aP_0^{-1}=(0.06\pm0.02)c$. We expect an energy shift of $\Delta E=E_{{\rm Fe K}\alpha}\Delta v/c=0.38\pm0.02$ keV, assuming an inclination angle $\iota=\pi/2$, typically adopted for Seyfert 2 sources, and assuming that the orbital plane of the black holes is aligned with the AGN inclination. This result is highly consistent with the observed centroid energies of the two Fe K$\alpha$ lines, found in \cite{severgnini18}  at $E_1=6.16\pm0.08$ and $E_2=6.56\pm0.15$ keV, i.e. $\Delta E=0.4\pm0.2$ keV. This strongly supports the hypothesis that the observed double-peaked iron line and periodicity are both produced by the same physical effect, likely the presence of two supermassive black holes in close orbit around each other. Hence, it is reasonable to consider our result (based on the {\it Swift}-BAT light curve) and the \cite{severgnini18} result (based mostly on the double Fe K line in {\it Swift}-XRT data) as two independent experiments of the same physical scenario; this allows us to combine the two p-values (see Sect.~\ref{sec:mcg}) and reject the null hypothesis at the combined confidence level of $2.9\sigma$. The results for Mrk 915 and MCG+11-11-032 are summarized in Table~\ref{tab:recap}.\\
\indent We note that the observed periodicity, in principle, may be produced by both the circumbinary disk periodic accretion flow into the cavity, or by the modulated emission by the minidisks, which has the same period as the binary. However, as argued by \citet{tang18}, the circumbinary emission is dominant in the soft X-rays ($E<4-5$ keV), while in the hard X-rays, including the BAT energy band, the observed emission is dominated by the minidisk emission. Moreover, concerning MCG+11-11-032, the consistency between the observed energy shift between the two Fe K$\alpha$ emission lines and the expected one inferred from the periodicity strongly favors the minidisk scenario. This means that the observed periodicity is likely the same as the binary system.\\
\indent Both Mrk 915 and MCG+11-11-032 would likely emit continuous gravitational waves with gravitational frequencies $f_{\rm gw}=2/P_0\simeq22$ nHz and $f_{\rm gw}\simeq30$ nHz, respectively, close to the frequency where PTAs are most sensitive \citep[e.g., ][]{lentati15,moore15,shannon15,mingarelli17,arzoumanian18}. However, current PTA data sets are not sensitive enough for any of these two sources, since GW signals from a single source can currently be observed only for sources with $\log(M/M_\odot)\gtrsim9.2$ at a distance of $d\lesssim150$ Mpc \citep{aggarwal19}. Therefore, the GWs emitted by both Mrk 915 and MCG+11-11-032 would be significantly below the current detection limit of PTAs.\\
\indent While waiting for the first PTA detection and the improvement of their sensitivity curve, X-rays are a very promising tool to unveil possible SMBHBs. Longer light curves are required for two main reasons: i) Confirm the current candidates with a higher significance. Note that the current 105-months {\it Swift}-BAT data release samples only approximately three periods for Mrk 915 and only nearly four for MCG+11-11-032. For the latter, the 123-months dataset has significantly improved the detection, from $2.6\sigma$ to $3.4\sigma$. The more data points we combine for each phase bin (see Sect.~\ref{sec:mrk915}), the smaller the errors on the epoch-folding will be, resulting in a much better $\chi^2$ test. ii) Find new candidates with the PSD selection criterium described in Sect.~\ref{sec:data}. In fact, by sampling longer continuous periods, while the noise level at a given frequency remains roughly constant, the periodic peak, if present, will be much higher at any additional sampled period. Longer {\it Swift}-BAT light curves, which will allow us to study the variability for up to $\sim250$ months, will therefore be crucial for periodicity studies. Additionally, the Wide Field Camera on board the future enhanced X-ray Timing and Polarimetry mission \citep[eXTP, ][]{zhang19} may continue the legacy of BAT even after the dismissal of the {\it Swift} telescope.\\
It must be stressed that the present work does not represent a full statistical study of the SMBHB population in the local Universe. First of all, it is limited to AGN, which are only $\sim10\%$ of the SMBHs in the observable Universe, and there is no reason to privilege active over non-active galaxies. Secondly, the selection criteria only selects the most prominent peaks and excludes sources with periodicities buried under a high level of noise. This is clear in the case of MCG+11-11-032, which was excluded from the PSD selection given its high level of non-periodic noise. This is also true for the method used in \citet{liu20}, in which a cut in excess variance has been applied, because the PSD does not take photometric errors into account, leaving both Mrk 915 and MCG+11-11-032 out of their sample. However, we note that we obtain the same results in the coincident sample, even if the two methods are different. Moreover, concerning MCG+11-11-032, we stress that we are able to reject the null hypothesis above $3\sigma$ only when we consider the 123-month long light curve, which samples around five periods.\\
\indent Finally, the detection of double-peaked Fe K$\alpha$ lines is still in its pioneering age. XMM-Newton, currently the instrument with the highest effective area in the Fe K$\alpha$ spectral region, is only able to observe energy shifts $\Delta E\gtrsim0.15$ keV. Therefore, double iron lines produced by orbital motions of SMBHBs with the range of periods analyzed in this work are only observable for a limited number of AGN, mostly with edge-on inclinations. Future microcalorimeters such as the ones on board XRISM \citep[e.g.,][]{xrism20} and Athena \citep[e.g.,][]{barret16} will revolutionize the search for binary SMBHs, being able to detect energy shifts down to $\Delta E\sim0.01$ keV, possibly opening the window for the search for double-peaked Fe K$\alpha$ lines in many more Seyfert galaxies and quasars.



\acknowledgments We thank the referee for useful comments that improved the quality of this paper. We thank Stefano Andreon, Alessandro Caccianiga, Sergio Campana, Sergio Frasca, Cristiano Palomba, Sara Rastello, Paolo Saracco and Fausto Vagnetti for discussions and suggestions. The authors acknowledge financial contribution from the agreements ASI-INAF n.2017-14-H.0 and n.I/037/12/0. CC acknowledges funding from the European Union's Horizon 2020 research and innovation programme under the Marie Sklodowska-Curie grant agreement No 664931. AS is supported by the European Research Council through the CoG grant ’B Massive’, grant number 818691. We acknowledge the use of public data from the {\it Swift} data archive.

\bibliographystyle{likeapj}
\bibliography{biblio}    

    
\end{document}